\documentclass[prd,onecolumn,nofootinbib,preprintnumbers,amssymb,amsfonts,amsmath,superscriptaddress,showpacs,hyperref]{revtex4-1}

\usepackage{graphicx}
\usepackage{bm}
\usepackage{color}
\usepackage{url}
\usepackage{epstopdf}
\usepackage{amsthm,mathrsfs} 

\newcommand{\be}{\begin{equation}}
\newcommand{\ee}{\end{equation}}
\newcommand{\baln}{\begin{align}}
\newcommand{\ealn}{\end{align}}
\newcommand{\ben}{\begin{equation*}}
\newcommand{\een}{\end{equation*}}

\long\def\symbolfootnote[#1]#2{\begingroup%
\def\thefootnote{\fnsymbol{footnote}}\footnote[#1]{#2}\endgroup}

\newcommand{\fr}{\frac}
\newcommand{\del}{\partial}

\newcommand{\Sc}{\mathcal{S}}

\newcommand{\x}{\boldsymbol{x}}

\begin{document}

\title{Tests of Quantum Gravity-Induced Non-Locality\\ via Opto-mechanical Experiments}
\author{Alessio Belenchia} 
\email{abelen@sissa.it}
\affiliation{SISSA, Via Bonomea 265, I-34136 Trieste, Italy and INFN, Sez. di Trieste.}
\affiliation{CNR-Istituto Nazionale di Ottica, Largo E. Fermi 6, I-50125 Firenze, Italy.}

\author{Dionigi M. T. Benincasa}
\email{dionigi.benincasa@sissa.it}

\author{Stefano Liberati}
\email{liberati@sissa.it}
\affiliation{SISSA, Via Bonomea 265, I-34136 Trieste, Italy and INFN, Sez. di Trieste.}

\author{Francesco Marin}
\email{marin@fi.infn.it}
\affiliation{Dipartimento di Fisica e Astronomia, University of Florence, Via Sansone 1, I-50019 Sesto Fiorentino, Firenze, Italy and INFN, Sez. di Firenze.}
\affiliation{European Laboratory for Non-Linear Spectroscopy (LENS), Via Carrara 1, I-50019 Sesto Fiorentino, Firenze, Italy.}

\author{Francesco Marino}
\email{marino@fi.infn.it}
\affiliation{CNR-Istituto Nazionale di Ottica, Largo E. Fermi 6, I-50125 Firenze, Italy and INFN, Sez. di Firenze.}

\author{Antonello Ortolan}
\email{antonello.ortolan@lnl.infn.it}
\affiliation{INFN, Laboratori Nazionali di Legnaro, Viale dell'Universit\`a, 2, 35020 Legnaro, Padova, Italy.}

\begin{abstract}
The 
non-relativistic limit of non-local modifications to the 
Klein-Gordon operator is studied, 
and the experimental possibilities
of casting stringent constraints on the non-locality scale 
via planned and/or current opto-mechanical experiments are discussed. 
Details of the perturbative analysis and semi-analitical simulations leading to the 
dynamical evolution of a quantum harmonic oscillator in the presence of non locality 
reported in~\cite{Belenchia:2015ake},
together with a comprehensive account of the experimental methodology 
with particular regard to sensitivity limitations related to thermal decoherence time and active cooling of the oscillator, are 
given. 
Finally, 
a strategy for detecting non-locality scales of the order of $10^{-22} \div 10^{-26}$ m
by means of the spontaneous time periodic squeezing of 
quantum coherent states is provided.
\end{abstract}

\maketitle

\section{Introduction}


Quantum gravity phenomenology is the phrase commonly used to describe the field of research 
that attempts to build a bridge between Planck scale theories of quantum gravity (QG) and observation.
The real challenge faced by the community working in this field is to derive phenomenology that is 
relevant at scales much lower than
the Planck scale, $M_p=1.22\times 10^{28}$ eV, where QG effects are expected to dominate, so that
existing models can be put to the test.
Over the last two decades there has been a steady stream of work in this direction. In particular,
relevant studies include: tests of quantum decoherence and state collapse models~\cite{Mavromatos:2004sz}, 
QG imprints on initial cosmological perturbations \cite{Weinberg:2005vy}, cosmological variation of coupling constants,
\cite{Damour:1994zq,Barrow:1997qh}, TeV Black Holes within extra-dimensions \cite{Bleicher:2001kh}, 
Planck-scale spacetime fuzziness \cite{AmelinoCamelia:1998ax}, generalised uncertainty principles 
\cite{Garay:1994en, Hossenfelder:2012jw, Marin:2013pga},  violations of discrete symmetries \cite{Kostelecky:2003fs} 
and violations of space-time symmetries \cite{Mattingly:2005re,Liberati:2013xla}. 
In this paper we add to this list by considering the phenomenological effects of a fundamental 
``spacetime nonlocality" in nonrelativistic, macroscopic quantum systems.


The underlying idea here is that models of QG with fundamental Lorentz invariance (LI)
lead to low energy effective theories with dynamics that are nonlocal in spacetime, 
once the high energy degrees of freedom have been integrated out.
Particular examples of this kind exist in causal set theory, 
where the interplay between Lorentz invariance and discreteness leads
to nonlocal dynamics for fields living on the causal set~\cite{Sorkin:2007qi};  string theory and string field 
theory where the string and its interactions are inherently nonlocal~\cite{Eliezer:1989cr}; 
and noncommutative geometry~\cite{Szabo:2001kg}. 
\footnote{It has also been argued that the same form of nonlocality must also be present in loop quantum gravity 
if it has any hope of preserving LI~\cite{Gambini:2014kba}.}
It appears therefore that, in general, theories of QG in which continuum spacetime 
emerges from more fundamental constituents and where LI is preserved, 
can only be realised at the expense of modifying low energy effective dynamics in 
an essentially nonlocal way. 

To be more specific let us consider a free massive scalar field, $\phi$, on a flat spacetime that has 
``emerged" from a LI theory of QG. The most na\"ive thing that one can imagine is that the
emerging field theory is just a standard local field theory described by the equations of
motion $(\Box+m^2)\phi(x)=0$.
A little more thought however reveals that this is unlikely to be the case. Indeed
any theory of QG gravity must at the very least contain
the scale $l_p=1.62\times10^{-35}$~m 
and therefore, following the usual ideas of effective field theory (EFT), it is natural to expect this 
scale to enter the low energy physics as a perturbative parameter in an expansion
around the local theories we know and love.
Thus, combining this with the fact that the theory is fundamentally LI,
the most natural dynamics that one might write down for such a field theory is something like
$f(\Box+m^2)\phi=0$, where $f$ is some non-polynomial function of the Klein-Gordon (KG) operator
such that $f(\Box+m^2)\rightarrow \Box+m^2$ in the limit $l_p\rightarrow 0$. 
\footnote{The infinite number of derivatives is crucial in order to avoid 
Ostrogradski's theorem~\cite{ostrogradski1850member}, which also applies to theories with
higher order, but finite, powers of the d'Alembertian operator $\Box$.}
In a sense one can think of $f$ as providing the UV completion of the 
EFT.

It should come as no surprise then that this is precisly what one find in the models referred to above.
For example, in four dimensions string field theory predicts a nonlocal
KG equation of the form~\cite{Koshelev:2011gx}
\begin{equation}
f(\Box+m^2)=(\Box+m^2)\,\exp{[l_p^2(\Box+m^2)]}\,,
\label{KGSFT}
\end{equation} 
while causal set theory gives~\footnote{Note that 
$f(\Box+m^2)$ was rigorously derived from causet theory only in the 
case $m=0$, but see discussion in \cite{Belenchia:2014fda} on ways of extending this
to the massive case.}
\begin{align}
f(\Box+m^2)=(\Box+m^2)
-\frac{3l_p^2}{2\pi\sqrt{6}}(\Box+m^2)^{2}\left[3\gamma-2+\ln\left(\frac{3l_p^2(\Box+m^2)^{2}}{2\pi}\right)\right]+\dots,
\label{4dEx}
\end{align}
where $\gamma$ is Euler-Mascheroni's constant. 
Note that in the first instance the function $f$ is an {\em analytic} function while in the second
it is {\em non-analitic}. 
Further it turns out that the scale 
entering the definition of $f$ need not be identified with the Planck scale itself
in general. This happens for example with causal set d'Alembertians, 
where theoretical considerations have led Sorkin to postulate that the scale
entering their definition is some $l_k\gg l_p$~\cite{Sorkin:2007qi}.
This is crucial for phenomenology since there is little hope in detecting nonlocal effects 
if they only become relevant at the Planck scale.
From here on we will therefore take the nonlocality scale $l_k$
to be a free parameter of the theory. 

%
%
%
%
%
%

In the rest of this paper we will explore a new phenomenological approach 
based on the application of the non-relativistic limit of an analytic nonlocal KG equation (e.g. \eqref{KGSFT}) 
to opto-mechanical quantum oscillators.
\footnote{We will not discuss the phenomenology of nonanalytic nonlocal QFTs here, but for recent
ideas on this we refer the reader to~\cite{Belenchia:2016sym}} 
We will argue that the true evolution of this system is governed by a nonlocal Schr\"odinger equation
whose specific form depends on the underlying nonlocal relativistic QFT. Finally we will 
analyse in perturbation theory the effects induced by the lowest order corrections to the standard Schr\"odinger
evolution. 
  
The paper is organised as follows. In Section~\ref{sec:nloc-nrel} we discuss the nonrelativistic limit
of nonlocal relativistic QFTs charachterised by analytic form factors $f$.
In particular, we will discuss the properties that a nonlocal QFT must possess in order
for there to exist a sensible physical interpretation of its nonrelativistic limit.  
Section~\ref{pert} describes the perturbative analysis of the nonlocal Schr\"odinger equation
in the presence of an external potential. 
In Section~\ref{nloc-Sch} we apply this analysis to the specific case of a harmonic oscillator potential, thus
reproducing the results reported in~\cite{Belenchia:2015ake} with a greater level of detail. 
Finally in Section~\ref{sec:exp} we discuss in detail the experimental strategies that can be 
used to cast limits on the non-locality scale with current, and near future, experiments involving
macroscopic quantum oscillators. Conclusions and a discussion of future work are given in 
Section~\ref{sec:concl}.

\section{Non--Relativistic Limit of Non-local Relativistic QFTs}
\label{sec:nloc-nrel}

Consider a free complex, massive, scalar nonlocal QFT defined by the Lagrangian
\be
\mathcal{L} = \phi(x)^*f(\Box+\mu^2)\phi(x)+c.c.,
\label{lag}
\ee
where $\Box = c^{-2}\del^2_t -\nabla^2$ and $\mu=mc/\hbar$. 
In order for the theory to be physically sensible we assume that the following conditions hold:
\begin{enumerate}
\item $f(k^2)=0$ iff $k^2=0$: this property ensures that there exist no classical runaway solutions 
and, when $f$ is entire, no ghosts. 
\item the nonlocal QFT must be unitary: conservation of probability.
\item the nonlocal QFT must possess a global $U(1)$ symmetry: this condition ensures that (some form 
of) a probabilistic interpretation can be given to the wave function.
\end{enumerate}
As already mentioned the function $f$ can be both entire analytic and non-analytic.
For the remainder of this paper we will assume that $f$ is entire analytic so that it can be expanded as
\be
f(z) = \sum_{n=1}^\infty b_n z^n.
\ee
Implicit in the definition of $f$ is the non-locality
scale $l_k$ which, in the local limit $l_k\rightarrow0$, sends $f(\Box+\mu^2)\rightarrow \Box+\mu^2$.
In particular we have that $b_n\propto l_{k}^{2n-2}$ and $b_1=1$. 

Following standard treatments (see e.g. Section 2.8 of~\cite{tong2006lectures})
we decompose the field as $\phi(x)=e^{-i \fr{m c^2}{\hbar}t}\psi(t,\x)$. Substituting this into our 
Lagrangian and taking the limit $c\rightarrow\infty$ we find
\be
\mathcal{L}_{\text{NR}}  = \psi^*(t,\x)f(\Sc')\psi(t,\x)+c.c.,
\label{lag2}
\ee
where NR stands for non-relativistic, $\Sc' =-\fr{2m}{\hbar^2}\Sc$, and
\be
\Sc=i\hbar \fr{\del}{\del t} + \fr{\hbar^2}{2m}\nabla^2
\ee
is the Schr\"odinger operator.

To derive the equations of motion we use a nonlocal generalisation of the Euler-Lagrange equations
\cite{bollini1987lagrangian} which gives 
\be
f(\Sc)\psi(t,\x)=0,
\label{nls}
\ee
where 
\be\label{oppb}
f(\Sc)\equiv\frac{-\hbar^{2}}{2m}f(\Sc')=\Sc+\sum_{n=2}^{\infty}b_{n}\left(\frac{-2m}{\hbar^{2}}\right)^{n-1}\Sc^{n}.
\ee
One can also include an external potential, $V(x)$, by adding the term $V(x)\psi^*\psi$ to the Lagrangian 
\eqref{lag2}.
To simplify notation we set $b_n =l_k^{2n-2}a_n$ so that equation~\eqref{oppb} becomes
\be
f(\Sc) = \sum_{n=1}^\infty  (-2m/\hbar^2)^{n-1} a_n l_k^{2n-2}\Sc^n.
\label{f}
\ee

The Lagrangian (\ref{lag2}) possesses a global $U(1)$
symmetry $\psi\rightarrow e^{i\alpha}\psi$ whose conserved current, $j_{NL}^\mu$, 
can be shown to be given by
\begin{align}\label{cur}
j_{NL}^0&=a_1\psi^*\psi -   i a_2 l_{k}^{2} \fr{2m}{\hbar}\psi^* \overset{\leftrightarrow}{\del_t}\psi 
-a_2 l_{k}^{2}\psi^*\nabla^2\psi-a_2 l_{k}^{2}\psi\nabla^2\psi^*+O(l_k^4)\\ 
j_{NL}^i&= -ia_1\fr{\hbar}{2m}\psi^*\overset{\leftrightarrow}{\nabla}\psi 
+ ia_2 l_{k}^{2}\fr{\hbar}{2m}\psi^*\overset{\leftrightarrow}{\nabla^3}\psi
+2a_2 l_{k}^{2}\dot{\psi^*}\overset{\leftrightarrow}{\nabla}\psi+O(l_k^4),
\end{align}
where $f\overset{\leftrightarrow}{\nabla^n}g=\sum_{i=0}^n (-)^i\nabla^if\nabla^{n-i}g$. 
Note that, as required by consistency with the local theory, 
\be
(j^0_{NL},j^i_{NL})\rightarrow (\psi^*\psi, -ia_1\fr{\hbar}{2m}\psi^*\overset{\leftrightarrow}{\nabla}\psi)
\ee
as $l_k\rightarrow 0$.

What we have so far is a nonrelativistic field $\psi$ satisfying a nonlocal generalisation of the 
Schr\"odinger equation. What we want though
is to be able to interpret $\psi$ as the wavefunction of a quantum mechanical system. The canonical way of doing
this in the local theory is to define a one-particle wavefunction for a generic one particle state constructed
from the field operator $\psi$, and show that this wavefunction satisfies the same Schr\"odinger equation as the field.
This analysis requires the Hamiltonian, which we currently lack in our nonlocal theory. Thus, from
here on we will proceed with the caveat that our model is only phenomenological in the sense that
we have yet to demonstrate that the one particle wavefunction of the nonrelativistic field satisfies the nonlocal
Schr\"odinger equation. We will comment more on this in the conclusions.

\section{Perturbative Analysis}
\label{pert}

Having laid down the foundations for a non-local Schr\"odinger evolution of a single particle quantum system,
we now turn to the problem of solving the nonlocal differential equation in the presence of a   
time independent potential $V(x)$. 

We wish to solve the nonlocal equation 
\be\label{nlse}
f(\Sc)\psi(t,x) = V(x)\psi(t,x),
\ee
where $f(\Sc)$ is some analytic function as in \eqref{f}, and $V(x)$ some physically reasonable binding potential.
Since the above equation is extremely hard to solve exactly for a given nontrivial potential $V$ 
we will solve it {\em perturbatively}.

In order to cast \eqref{nlse} in a form amenable to a perturbative analysis, we first note that the presence 
of an external binding potential $V(x)$ introduces an energy scale that can be parametrized as 
$\hbar \omega$, where the ``scale'' $\omega$ has dimensions of (time)$^{-1}$. 
We can use this new scale together with the other scales in the problem to construct a dimensionless 
parameter $\epsilon:=m\omega l_k^2/\hbar$. 
For physically reasonable choices $m$, $\omega$ and $l_k$, $\epsilon$ is much smaller than unity so that we
can use it as our perturbative parameter in the expansion of $f(\Sc)$:
\be
f(\Sc)= \Sc-\fr{2a_2}{\hbar\omega}\epsilon\, \Sc^2+\sum_{n=3}^{\infty}a_{n}\left(\frac{-2}{\hbar\omega}\right)^{n-1}\epsilon^{n-1}\Sc^{n}.
\label{pertsch}
\ee  
Next we will assume that \eqref{pertsch} admits solutions of the form
\be
\psi= \sum_{n=0}^{\infty}\epsilon^{n}\psi_{n}.
\label{expansion}
\ee
Substituting \eqref{expansion} into \eqref{pertsch} we find the following set of differential equations
\begin{align}
O(1):&\quad (\Sc-V)\psi_0=0,\\
O(\epsilon):&\quad (\Sc-V)\psi_1=J_1\label{j1},\\
O(\epsilon^2):&\quad (\Sc-V)\psi_2=J_2\label{j2},\\
&\quad\;\text{etc.}
\nonumber
\end{align}
where the $J_i$, $i=1,2,\dots$ are source terms. Note that the $i$-th source term 
depends on the solution to the $(i-1)$th order problem, for example
\begin{align}
J_1&=\fr{2a_2}{\hbar\omega}\Sc^2\psi_0\label{source1},\\
J_2&=\fr{-4a_3}{\hbar^2\omega^2}\Sc^3\psi_1,\quad \text{etc.}
\end{align}

Implicit in the above analysis is the assumption that $\psi_0$ -- a solution to the standard Schr\"odinger equation -- 
is also an approximate solution to the nonlocal equation, i.e. that
\be
|(f(\mathcal{S})-V)\psi_0|= O(\epsilon)\ll1,\quad\forall t,\;x.
\label{condition}
\ee
The idea behind this assumption is that nonlocal extensions, $f$, of experimentally verified
local models must be such that they admit solutions to the local models as approximate solutions.  
Clearly this assumption is difficult to check explicitly, especially here where we have a 
function of the operator $\Sc$ containing both space and time derivatives. 


We can summarise our perturbative approach as follows:
\begin{itemize}
\item Consider nonlocal Schr\"odinger equations with entire analytic $f(\Sc)$s
in the presence of an external potential $V(x)$ that satisfy \eqref{condition}.
\item Using the scale introduced by the potential construct a (small) dimensionless parameter $\epsilon$.
\item Expand $f(\Sc)$ in $\epsilon$ and assume that solutions can be written as \eqref{expansion}.
\item Solve the problem order by order in $\epsilon$ checking that the conditions are satisfied at each order.
\item Finally, one should check for consistency that each term $\epsilon^n\psi^n$ is indeed smaller than the 
previous one $\epsilon^{n-1}\psi^{n-1}$ for each $n$ (up to the relevant order of interest). 
\end{itemize}

\section{Nonlocal Schr\"odinger Equation in (1+1)D with a Harmonic Oscillator Potential}
\label{nloc-Sch}

Consider a single particle in a harmonic oscillator potential in 1+1 dimensions satisfying the
equation
\be\label{nlse2}
f(\Sc)\psi(t,x) = \fr{1}{2}m\omega^{2} x^2\psi(t,x),
\ee
where $m$ is the mass of the system and $\omega$ its
angular frequency. Following the steps laid out in the previous section
we construct the dimensionless parameter $\epsilon\equiv m\omega l_k^2/\hbar$
and write $f(\Sc)$ as:
\be
\left(\Sc-\fr{2a_2}{\hbar\omega}\epsilon\, \Sc^2
+\sum_{n=3}^{\infty}a_{n}\left(\frac{-2}{\hbar\omega}\right)^{n-1}\epsilon^{n-1}\Sc^{n}\right)
\psi(t,x)= \fr{1}{2}m\omega x^2\psi(t,x).
\label{schrosc}
\ee
In order to keep the notation as clear as possible we define the following dimensionless variables
$\hat{t}=\omega t$, $\hat{x}= \sqrt{\omega m/\hbar}\,x$, and  $\hat{\psi}=\gamma \psi$, 
where $\gamma$ has dimensions of $1/\sqrt{\rm Length}$, so that \eqref{schrosc} becomes
\be\label{seqn}
\left(\hat{\mathcal{S}}-2a_2\epsilon\hat{\mathcal{S}}^2
+\sum_{n=3}^{\infty}a_{n}\epsilon^{n-1}(-2)^{n-1}\hat{\mathcal{S}}^{n}\right)\hat{\psi}=\frac{1}{2}\hat{x}^{2}\hat{\psi}.
\ee 
Throughout the rest of this section we will use these dimensionless variables but will drop the 
hat symbol for notational simplicity. 

We assume that \eqref{schrosc} admits solutions of the form \eqref{expansion}. In particular,
we will be interested in solutions that are perturbations around the coherent state
\be
\psi_0:=\psi_{\alpha}(t,x)=\frac{1}{\pi^{1/4} } \ \exp\left[\sqrt{2} \alpha  
e^{-i t} x-\frac{1}{2} \alpha ^2 e^{-2 i t}-\frac{\alpha ^2}{2}-\frac{i t}{2}-\frac{x^2}{2}\right],
\label{coh}
\ee 
where without loss of generality $\alpha$ can be taken real and $(\Sc-x^2/2)\psi_0=0$.
This choice of $\psi_0$ is motivated by the fact that coherent states are 
relatively easy to realise within the experimental setting we have in mind (see Section~\ref{sec:exp})
and furthermore include the harmonic oscillator's ground state as a  specific case. 

Next we want to solve the differential equation at order $\epsilon$. To this end we first substitute $\psi_0$
into \eqref{source1} to find
\be
J_1=a_{2}\frac{1}{2\pi^{1/4}}e^{-\frac{1}{2}e^{-2it}\left(-2\sqrt{2}e^{it}x\alpha+\alpha^{2}+e^{2it}(3it+x^{2}+\alpha^{2})\right)}\left(e^{it}(2-4x^{2}+x^{4})+4\sqrt{2}x\alpha\right).
\ee
Then, to solve $(\Sc-V)\psi_1=J_1$ we use the ansatz
\be
\psi_{1}(t,x)= \psi_{0}(t,x)\left[c_{0}(t)+c_{1}(t) x+c_{2}(t)x^2 +c_{3}(t)x^3 +c_{4}(t)x^4 \right].
\label{eqanz}
\ee 
which leads to the following system of ordinary differential equations for the time dependent 
coefficients $c_{i}(t)$:
\begin{align}\label{syst}
0&=2 i \dot{c}_{4}(t)-8 c_{4}(t)-a_{2},\nonumber\\
0&= i e^{i t} \dot{c}_{3}(t)-3 e^{i t} c_{3}(t)+4 \sqrt{2} \alpha  c_{4}(t), \nonumber\\ 
0&= e^{i t} \left(i \dot{c}_{2}(t)+6 c_{4}(t)+2 a_{2}\right)-2 e^{i t} c_{2}(t)+3 \sqrt{2} \alpha  c_{3}(t), \\ 
0&= \sqrt{2} \alpha  c_{1}(t)+e^{i t} \left(i \dot{c}_{0}(t)+c_{2}(t)-a_{2}\right),
\nonumber\\ 
0&= 4 i e^{7 i t} \dot{c}_{1}(t)-4 e^{7 i t} c_{1}(t)-\sqrt{2} \alpha  a_{2} \left(-3 \alpha ^2+6 \left(2 \alpha ^2+1\right) e^{2 i t}
+\,e^{4 i t} \left(-4+\alpha ^2 (-9+12 i t)\right)+6 e^{6 i t}\right),\nonumber
\end{align}
which we solve using Mathematica 11 subject to the initial condition $\psi_1(0,x)=0$.
Solutions to \eqref{syst} with the given initial condition contain secular terms which grow linearly in time as $\epsilon t$. 
These terms are a well known artefact due to the non uniform convergence of the perturbative expansion.  
To avoid the appearance of secular terms we used the method of multiple scales and
refer the reader to Appendix~\ref{app:multiple} for further details. 

Finally we find 
\begin{align}\label{eq:coef1}
    & c_{0}(t)=\frac{1}{32} a_{2} \,e^{-8 i t} \left(\alpha ^4-8 \alpha ^4 e^{2 i t}+8 \alpha ^4 e^{6 i t}-\alpha ^4 e^{8 i t}-6 \alpha ^2 e^{2 i t}+20 \alpha ^2 e^{4 i t}\right.\\ \nonumber
    &\quad\quad\quad\qquad\qquad\quad \left. -14 \alpha ^2 e^{6 i t}+28 \alpha ^2 e^{8 i t}-3 e^{4 i t}-4 e^{6 i t}+7 e^{8 i t}\right),\\
    & c_{1}(t)=-\frac{1}{4 \sqrt{2}} \alpha\,  a_{2}\, e^{-7 i t} \left(\alpha ^2-6 \alpha ^2 e^{2 i t}+3 \alpha ^2 e^{4 i t}+2 \alpha ^2 e^{6 i t}-3 e^{2 i t}+4 e^{4 i t}-e^{6 i t}\right),\\
    & c_{2}(t)=\frac{1}{8} \,a_{2}\, e^{-6 i t} \left(3 \alpha ^2-12 \alpha ^2 e^{2 i t}+9 \alpha ^2 e^{4 i t}-3 e^{2 i t}-2 e^{4 i t}+5 e^{6 i t}\right),\\
    & c_{3}(t)=-\frac{1}{2 \sqrt{2}}\, \alpha\,  a_{2}\, e^{-5 i t} \left(1-e^{2 i t}\right)^2,\\
    & c_{4}(t)=\frac{1}{8} a_{2}\, e^{-4 i t} \left(1-e^{4 i t}\right).
\end{align}
Although we do not show it here, the same procedure, including an ansatz similar to eq.~\eqref{eqanz} but with a polynomial of order 8, can be used to solve the nonlocal Schr\"odinger equation to 2nd order in $\epsilon$.

\subsection{Wave function normalisation}

With the first order perturbative solution to eq.~\eqref{schrosc} at hand, we can now compute expectation 
values and variances of physical observables in this state. 
However, for these to make sense we need to first ensure that a probabilistic interpretation
of the wavefunction exists. This requires at the very least that the following condition 
$\int_{-\infty}^\infty dx\,|\psi(x)|^2=1$ holds.
Now recall that the conserved charge in eq.~\eqref{cur} is not simply $|\psi|^2$, so that expectation values directly computed using 
$|\psi_0+\epsilon\psi_1|^2$ cannot have a well-defined probabilistic interpretation  at order $\epsilon$. 
A quick fix to this problem that leads to a well-defined conserved
probability distribution (at least to this order in $\epsilon$), 
is to normalise the wavefunction $\psi_{0}+\epsilon\psi_{1}$ 
using its own norm.
In accordance with the Born rule the probability density is then given by
\begin{equation}
\rho(t,x)= \frac{\psi^*(t,x)\psi(t,x) }{\int^\infty_{-\infty} |\psi|^2 {\rm d}x},
\label{prob}
\end{equation}
so that $\int_{-\infty}^\infty dx\,\rho(x)=1$ by construction and is therefore conserved. 
It should be noted that the normalisation factor is 1 at order $\epsilon$ when considering perturbations
around the ground state, i.e. $\langle\psi_{0}|\psi_{1}\rangle=0$, while in the case of a generic coherent 
state an order $\epsilon$ time dependent correction is present. 
The above normalisation factor ensures that even in this case we a have a meaningful probability distribution.

\subsection{Phenomenology}
Given the probability distribution \eqref{prob} we can compute the mean and variance of the position and momentum
of the particle. We find
\begin{align}
\langle x\rangle &=\sqrt{2} \alpha  \cos (t)\, \left(1+ \frac{ 1}{4} \epsilon  \alpha ^2 a_{2} \left[ \cos (2 t)-1 \right] \right)+\mathcal{O}(\epsilon^{2}),
\label{meanxcoherent}\\
\langle p\rangle &= \sqrt{2} \alpha  \sin (t)\left( 1+ \frac{1}{4}   \epsilon \,  a_2\  \left[ \alpha ^2 (7+3  \cos (2 t))-2 \right]\right) +\mathcal{O}(\epsilon^{2}),
\label{meanpcoherent}\\
\mbox{Var}(x)&=\frac{1}{2}\left( 1- \epsilon  a_2 \left[\left(6 \alpha ^2-1\right) \sin^2 ( t) \right]\right)+\mathcal{O}(\epsilon^{2}),\label{varxcoherent}\\
\mbox{Var}(p)&=\frac{1}{2}\left( 1+  \epsilon  a_2 \left[\left(6 \alpha ^2-1\right) \sin^2(t) )\right]\right)+\mathcal{O}(\epsilon^{2})\label{varpcoherent}.
\end{align}

\begin{figure}[htpb]
\centering
\includegraphics[width=0.5\textwidth]{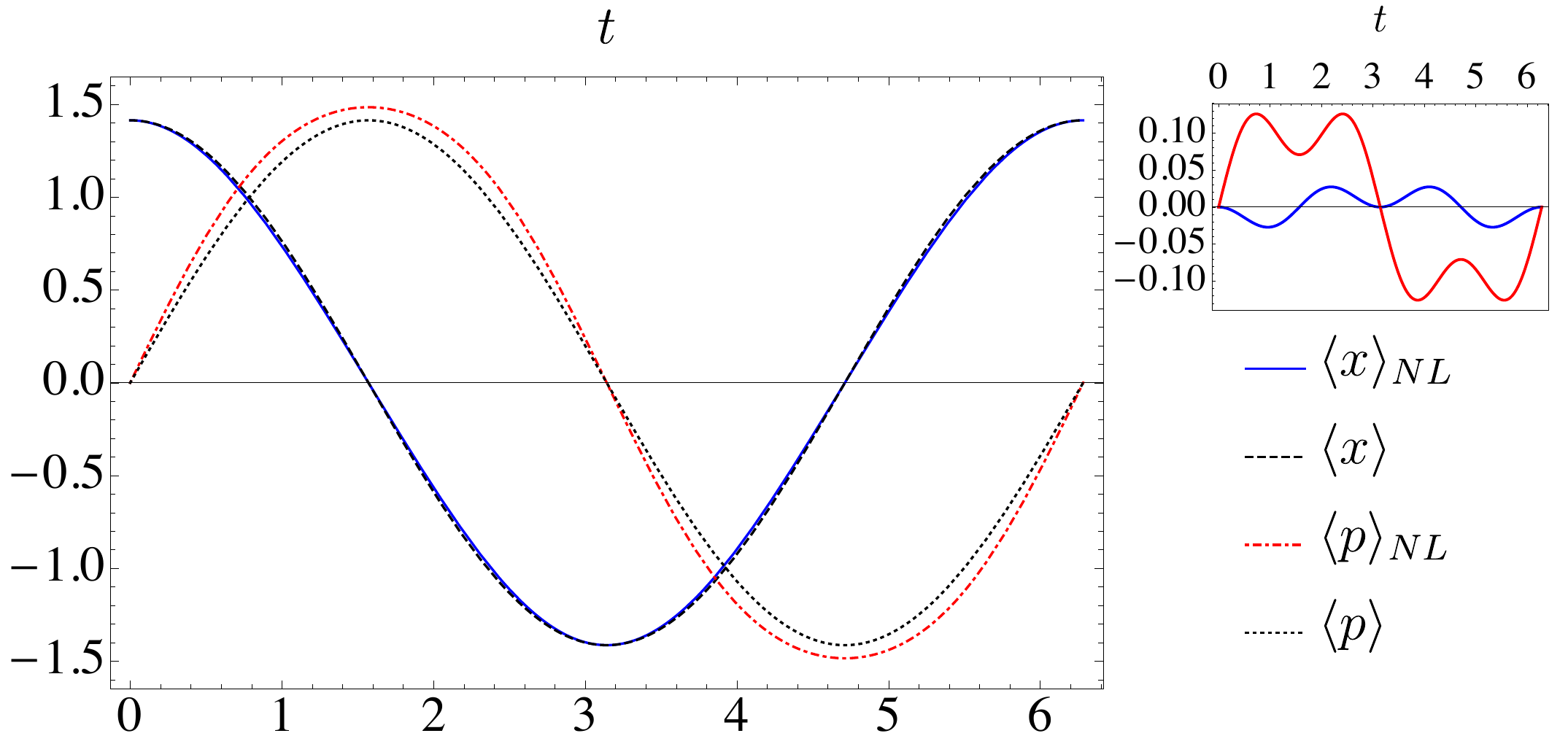} 
\caption{Periodic time dependence of the mean position and momentum for a coherent state (colours online) for
$\alpha=1$, $a_{2}=1$ and we have set $\epsilon=10^{-1}$ to amplify the effect of the nonlocality.
The continuous blue and red lines represent the mean position and momentum in 
eqs.~\eqref{meanxcoherent} and~\eqref{meanpcoherent} respectively. The black dotted and black dashed lines 
represent the mean position and momentum for the standard coherent state respectively. 
The insert on the top right shows the order $\epsilon$ corrections to 
the standard coherent state for the mean position (blue) and momentum (red).}\label{med}
\end{figure}

\begin{figure}[htpb]
\centering
    \includegraphics[width=0.5\textwidth]{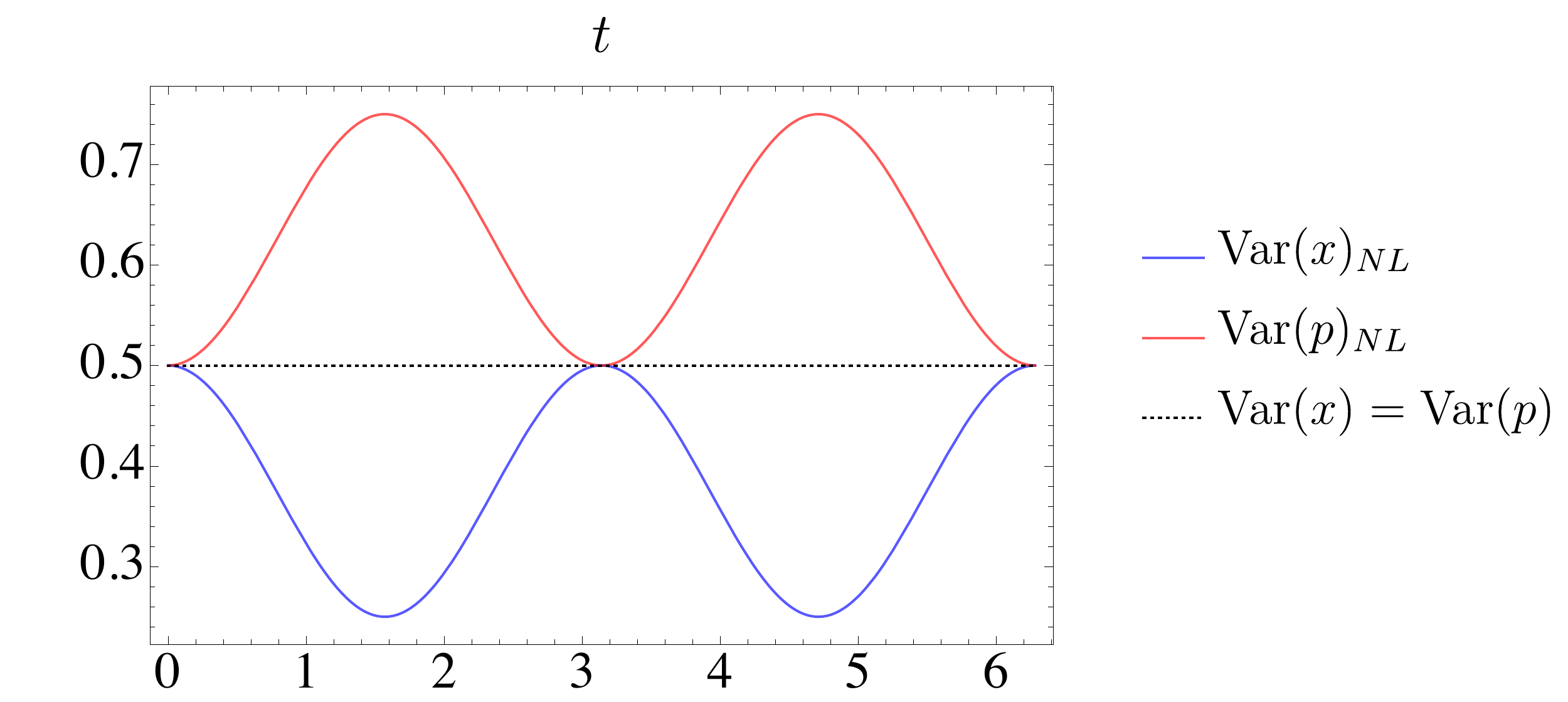} 
\caption{Periodic time dependence of the variances of position and momentum for a coherent state (color online) for $\alpha=1$, $a_{2}=1$ and $\epsilon=10^{-1}$. The continuous blue and red lines represent the variance of  position and momentum in eqs.~\eqref{varxcoherent} and~\eqref{varpcoherent} respectively. The black dotted line represents the variance of position and momentum for the standard coherent state which is equal to $1/2$. }\label{var}
\end{figure}

So, on the basis of Eqs.~\eqref{meanxcoherent}-\eqref{varpcoherent}, the effects of nonlocality 
appear in the form of deviations from the standard variances and mean values of position 
and momentum, as shown in Figs.~\ref{med},\ref{var}. 
In particular, our model predicts an oscillatory 
behaviour of the variance of $x$, together with a time-averaged expectation value that is larger than 
the standard $x_{\rm{zpm}}^2 = \hbar /2 m \omega$, and a third-harmonic distortion in the evolution 
of coherent states. The strength of these effects is governed by the perturbation parameter $\epsilon$.

Let us remark that $\sqrt{\epsilon}$ is given by the ratio between $l_k$ and the size of the ground-state 
wavepacket $x_{\rm{zpm}} = \sqrt{\hbar /2 m \omega}$, 
i.e.~the zero-point fluctuations. Such dependence suggests that massive 
quantum systems or, more precisely, systems with the smallest zero-point fluctuations, could be the ideal 
setting for detecting such nonlocality.
Furthermore, $\mbox{Var}(x)\mbox{Var}(p)=1/4 +O(\epsilon^2)$ so that the perturbed state is still a state
of minimum uncertainty but one which undergoes a spontaneous, cyclic, time dependent ``squeezing" in 
position and momenta (see Figure~\ref{var}). The word ``squeezing" is apt in view of the fact that the state is one of minimum uncertainty.

Finally, it is worth noting that the expectation values of $x$ and $p$ in the ground state ($\alpha=0$) are 
identical to the standard local case to first order in $\epsilon$ (and indeed the same holds true to second order);
while the variances are always modified to order $\epsilon$, except for the peculiar case $\alpha=\pm 1/\sqrt{6}$. 
This peculiarity appears to be a numerical accident --- as is confirmed by going to order $\epsilon^2$ where the
values $\alpha=\pm 1/\sqrt{6}$ play no special role --- so we attach no particular physical meaning to it.

\subsection{Range of validity of the perturbative expansion}

Before we proceed, an important point to make is that the validity of the perturbative expansion 
depends on the state that we choose to
expand around. In this case, expanding around the coherent state \eqref{coh} implies that the validity of the 
expansion will also depend $\alpha$. 
To see this consider the $L_2$-distance between $\psi_0$ and $\psi$ to first order in $\epsilon$:
$$
\|\psi_{0}-\left(\psi_{0}+\epsilon \psi_{1}\right)\|  \propto \epsilon |\alpha|^{4}\, .
$$
This tells us that for the perturbative expansion to be valid one must require
that $\epsilon\alpha^{4}\ll 1$, and not just $\epsilon\ll1$ . 

\begin{figure}[htpb]
\centering
    \includegraphics[width=.5\textwidth]{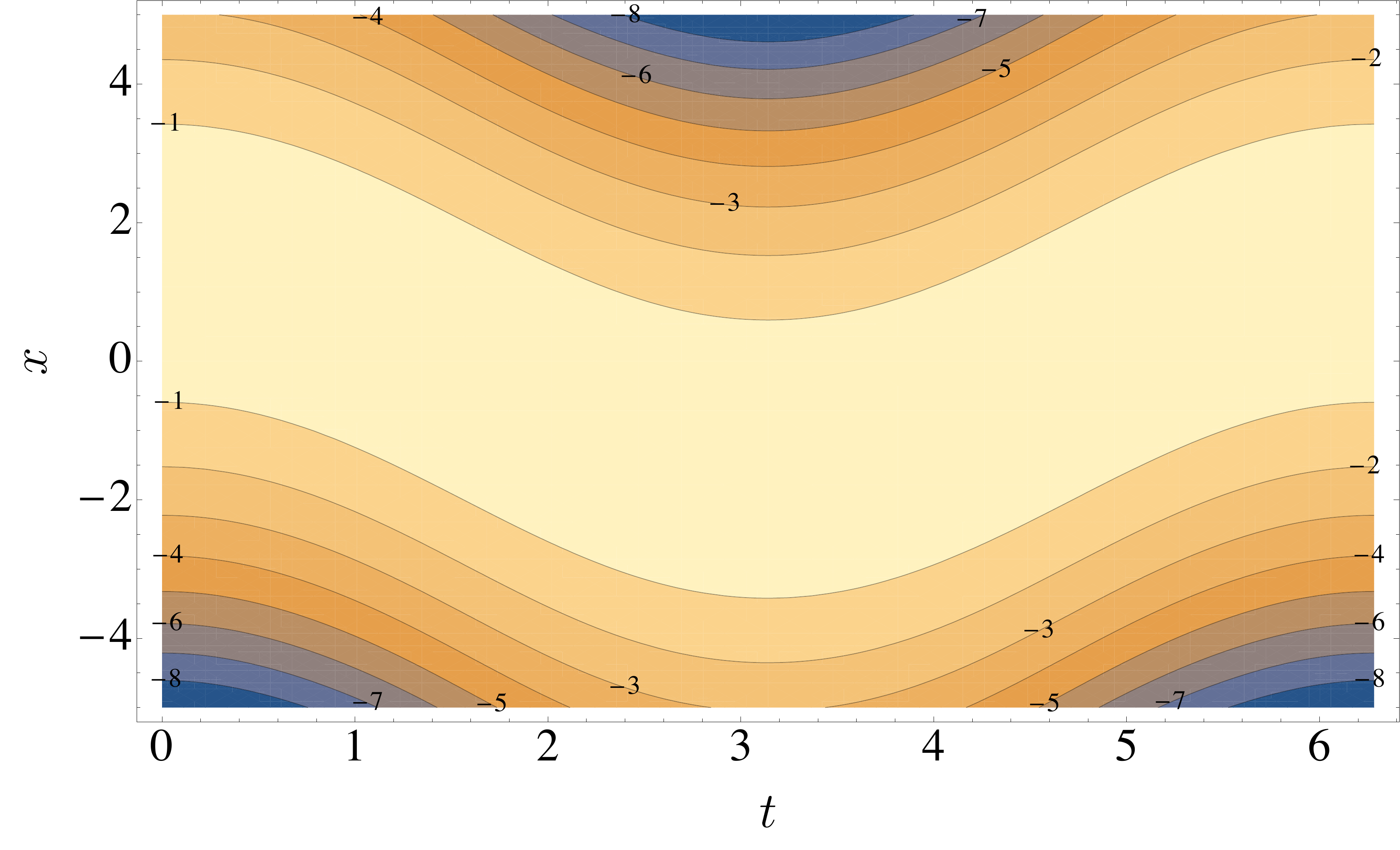} 
\caption{A contour plot of $\log_{10}(|\epsilon \psi_1|/|\psi_0|)$ for $\alpha=1$, $\epsilon=10^{-3}$ and $a_2=1$.
Note that although $\psi_{1}$  dominates over $\psi_{0}$ for large $|x|$ because of the form 
of our ansatz at spatial infinity, 
the total wavefunction is still suppressed by a Gaussian factor.
Furthermore, due to the presence of a binding potential,  these regions are irrelevant for the experimental system we are modelling. We have also checked the analogous condition for the order $\epsilon^{2}$ solution with similar results.}
\label{wavefn}
\end{figure}

Finally we checked for consistency that $|\epsilon \psi_1|/|\psi_0|\ll1$ (at least in the spacetime 
region relevant for the actual systems under consideration), with results shown in Figure~\ref{wavefn}.

\section{Optomechanical Tests of Nonlocality}
\label{sec:exp}

In this section we provide a detailed discussion of possible experimental tests of our model using opto-mechanical 
experiments based on macroscopic quantum oscillators. For definiteness we will assume that $a_2=1$. 

Nowadays optomechanical experiments can cool a macroscopic oscillator down to thermal occupation numbers 
below unity, as well as prepare mechanical squeezed states. In most cases the mechanical system 
is coupled to an electromagnetic field that is either used to prepare the oscillator in its quantum ground state or 
to monitor its motion. In these setups the wavefunction  $\psi$ is associated to an effective coordinate describing the displacement of a normal mode, or, 
to some approximation, to the centre-of-mass motion of the mechanical oscillator.

Let us stress, that the modelling of optomechanical interactions in the presence of nonlocality is not 
straightforward and currently is not included in our prototype model. Therefore, limits obtainable from 
current experiments, while providing preliminary hints on the length scales achievable in optomechanical 
setups, should not be used for a quantitative comparison with our model.
Nevertheless, it is still possible to conceive experimental schemes, based on state-of-the art technologies, 
that could potentially improve current limits on the nonlocality scale.

It is also worth mentioning that first bounds have already been obtained by comparing nonlocal 
relativistic EFTs to the 8 TeV LHC data \cite{Biswas}, in which the authors find $l_k \le 10^{-19}$ m. 
So it would be of great interest to realise independent experiments able to explore new intermediate 
regimes between the LHC and the Planck scale. In the following we provide first estimates of the limits 
achievable via optomechanical experiments.

\subsection{Limits from ground-state variance}

First constraints on nonlocal effects can be imposed by comparing the measured variance of $x$ 
with the corresponding predictions for the ground state. Taking the time average of Eq.~\eqref{varxcoherent} 
with $\alpha=$0 we find that $\text{Var}(x)$ is increased with respect to its standard value by $\epsilon/2$. 
In order to compare our predictions with experiments we require both an oscillator energy close enough 
to its standard ground value (namely, with an average occupation number $\langle n\rangle \ll 1$), and a 
sufficient accuracy, $\Delta_{\rm meas}$, in the measurement of the variance. With these conditions we derive 
from Eq.~\eqref{varxcoherent} an upper limit to $\epsilon$ of the form 
$\epsilon < 2 (\langle x^2\rangle _{\rm{meas}}/ x_{\rm{zpm}}^2-1) = 4 \Delta_{\rm{meas}}$ 
and thus a bound on the nonlocality scale, $l_k < 2 x_{\rm{zpm}} \sqrt{\Delta_{\rm{meas}}}$.

The cooling of a mechanical oscillator close to the quantum ground state can be achieved by means of ultra-cryogenic techniques, e.g. by dilution refrigerators, or by active radiation-pressure cooling starting from pre-cooled oscillators. In the first case, the oscillator is naturally in thermal equilibrium with the cryogenic environment, with temperaratures typically around a few tens of mK. In these conditions, an average occupation number $\langle n\rangle < 1$ can be obtained for mechanical oscillators with resonant frequencies in the GHz range, and thus with very low masses.

Recent experiments have cooled silicon optomechanical crystals, reaching an average phonon occupancy as low as $\langle n\rangle \approx 0.02$, which has been measured by single-phonon-counting techniques using weak optical excitation pulses \cite{meenehan,riedinger}. In these studies a measurement of the variance of $x$ is not provided, but it could be realised, in principle, by observing the phase fluctuations of the field reflected by the cavity in an interferometric setup. The measurement should be performed in a time shorter 
than the thermal decoherence time $\tau=(Q/\omega)/(1 + \langle n\rangle) \approx (Q/\omega)$. 
Here a crucial problem is to achieve the required sensitivity while using a sufficiently weak probe to avoid the quantum back-action of the measurement field. The latter imposes a futher limitation on the measurement time, which now should be shorter than $\tau_{BA}=(Q/\omega)/(1+\langle n_{BA} \rangle)$, where $n_{BA}$ is the number of thermal phonons produced by the back-action (back-action heating).

In the bad-cavity regime and with a probe power corresponding to the standard quantum limit, one basically obtains $\langle n_{BA} \rangle \approx$ 1. We thus consider a single measurement time $\tau_{\rm{meas}} \sim (Q/\omega)$ and an optical power smaller by a factor of ten (i.e. $\langle n_{BA} \rangle =$0.1). In these conditions, the signal-to-noise ratio achieved in a single measurement is just $0.1$, but the cycle can be repeated several times in order to reach an accuracy of the order of 1$\%$ that is comparable to the reasonably predictable systematic errors in the estimate of the system's parameters.

The experiments in Refs. \cite{meenehan,riedinger} use similar nanomechanical oscillators with resonant frequencies around 5 GHz. A direct measurement of the mass is not provided, but it can be roughly estimated from the dimensions of the moving part of the photonic crystal to be of the order of $m \approx 10^{-15}$ kg. The corresponding zero-point fluctuations are $x_{\rm{zpm}} \sim 3$ fm$=10^{-15}$ m. Assuming an accuracy $\Delta_{\rm{meas}} \sim 1\%$, we obtain $l_k < 6 \times 10^{-16}$~m.

We can also consider mechanical oscillators that are actively cooled by radiation pressure starting from cryogenic or ultra-cryogenic conditions. By means of this technique it is possible to achieve ground state cooling of oscillators with larger $m \omega $ and then with a lower $x_{\rm{zpm}}$. On the other hand, the oscillator is now kept in a dynamic thermal equilibrium with a hot background (corresponding to thermal occupancies $\langle n\rangle \gg 1$) and a cooling bath provided by the optomechanical interaction. As mentioned before, the effects of the radiation-pressure coupling between optical and mechanical degrees of freedom are neglected in our model. Therefore, we cannot provide any prediction on how nonlocal effects will be modified by such interaction. For a meaningful comparison with our model, the measurement of the variance should be realised within a time $\tau$, after turning off the cooling laser.

In order to assess the feasibility of such measurements, we consider two recent experiments with actively cooled oscillators: an aluminum membrane coupled to microwave radiation that is used to cool and monitor its motion \cite{lecocq} and a SiN membrane in a high-finesse optical cavity \cite{peterson}.
In the first case the system's parameters are $m=5 \times 10^{-14}$~kg, $\omega/2\pi= 15$ MHz, corresponding to $x_{\rm{zpm}} \simeq 5$ fm, and a quality factor $Q=1.6 \times 10^6$. The background temperature is $T = 30$ mK, corresponding to an occupation number $\langle n \rangle=42$, which is then reduced to $\langle n_{\rm{eff}} \rangle \sim 0.2$ by active cooling. In the second case, $m=9 \times 10^{-12}$ kg, $\omega/2\pi= 1.5$ MHz ($x_{\rm{zpm}} \simeq 1.2$ fm), $Q=8 \times 10^6$, $T=0.36$ K ($\langle n \rangle=4800$) and $\langle n_{\rm{eff}} \rangle \sim 0.2$. Using the above values of $\langle n \rangle$, we obtain respectively $\tau \simeq 0.4$ ms ($6000$ oscillation periods) and  $\tau \simeq 0.18$ ms ($270$ oscillation periods). Since the measurement time is much shorter than in the previous case, 
back-action can be neglected here.

If we perform a single measurement with sensitivity corresponding to the standard quantum limit for continuous detection, the signal-to-noise ratio achieved is $1/\langle n \rangle$ (the detection spectral bandwidth is $\langle n \rangle$ times the natural linewidth of the mechanical resonance). In order to achieve an accuracy of around $10 \%$, the measurement should thus be repeated $100 \langle n \rangle^2$ times, which can be reasonably done for the oscillator with $\langle n \rangle=42$. In this case we obtain $l_k < 2 \times 10^{-15}$~m. The assumption of a sensitivity kept at the standard quantum limit can be relaxed due to the short interaction time that limits the effect of the back-action. In principle, the sensitivity could even be increased by a factor close to $\langle n \rangle$, thus also making experiments with a larger $\langle n \rangle$ feasible. This can be accomplished, e.g., by increasing the measurement laser power. However a major improvement is technically challenging, so we keep our previous conservative assumption.

So far we have considered the limits achievable by existing optomechanical experiments, but 
it is interesting to explore potential further advances.
As we have shown, a small width of the ground-state position wavepacket, and thus a large product $m \omega$, is a favourable 
characteristic for the purpose of reaching lower limits on $l_k$. In general, 
the larger the frequency is, the lower the mass is, 
although a larger $m \omega$ is more easily achieved in massive oscillators, 
where the relatively low frequency is more 
than offset by the large modal mass. However,
experiments with low-frequency oscillators require lower temperatures to reach the quantum ground state, as well as higher sensitivities in position measurements due to the reduced $x_{\rm{zpm}}$. In this context, resonant gravitational bar-detectors represent the state-of-the-art, since they are 
designed to detect extremely small displacements and therefore exhibit very low background length fluctuations. 

For instance, the first longitudinal mode of the AURIGA detector (a 2.3 ton aluminium bar with the first longitudinal mode oscillating at 1kHz)
has been cooled down to the millikelvin regime using a cold damping technique \cite{vinante}. 
Due to its large mass and relatively low temperature, it displays rms position fluctuations as low as $\sim 6 \times 10^{-19}$m. The oscillator is in a thermal state and it should be further cooled to an effective temperature  five orders of magnitude smaller 
in order to approach the quantum ground state. Moreover, the detection system should be sensitive enough 
to measure the corresponding zero-point fluctuations at the level of $x_{\rm{zpm}} \sim 10^{-21}$~m, 
which is very far from being trivial to do.
We thus turn our attention back to micro-oscillators that may enter the quantum regime in the near future. 

For the purpose of this discussion we assume as reasonable experimental parameters  a mechanical frequency $\omega/2\pi= 300$ kHz (at frequencies below 100 kHz acoustic and/or technical noise are usually too strong), a mechanical quality factor $Q=10^7$ and a background displacement noise (i.e. the sensitivity) of $10^{-38}$ m$^2$/Hz. Starting from a base temperature of $0.1$K, the mechanical mode should be actively cooled to an effective quality factor $Q_{\rm{eff}}=1500$, in order to reach a thermal occupancy $\langle n_{\rm{eff}}\rangle =1$. In these conditions, setting the sensitivity as a lower limit to the final peak spectral density, we derive $m=10^{-5}$ kg and thus $x_{\rm{zpm}} \simeq 2 \times 10^{-18}$ m. On the basis of these considerations, masses of around $10^{-5}$ kg represent a reasonable limit for the achievement and measurement of the quantum regime in mechanical resonators.

Opto-mechanical devices with similar characteristics have already been realised, though not yet cooled down to their quantum ground state. For instance, the literature reports silicon micro-mirrors with flexural-torsional modes oscillating at $\omega/2 \pi = 100-200$~kHz, masses of $m=2-3 \times10^{-7}$~kg and mechanical quality factors, measured at cryogenic temperatures, of $Q=1-3 \times 10^6$ \cite{serra,borrielli} as well as quartz micropillars with a compression-dilatation mode at $\omega/2\pi = 3-4$ MHz, $m=2-4 \times 10^{-8}$~kg and $Q=2.5 \times 10^7$ \cite{Kuhn,Neuhaus}. For all these devices, the zero-point fluctuations are of the order of $x_{\rm{zpm}}\simeq10^{-2}$ fm.
Operating at a background temperature of $100$ mK, we have $\langle n\rangle \sim 600$ for the quartz oscillator, and a reasonable upper limit for $l_k$ of a few $10^{-18}$~m.

Summarizing, the bounds on the nonlocal scale obtainable with measurements on the ground state range from $l_k \le 10^{-15}$ m to $l_k \le 10^{-18}$ m. 
The latter is reasonably close to the constraint obtained at LHC.

\subsection{Evolution of coherent states}\label{evol}

A further comparison between our theory and experiments can be based on the evolution of coherent states. As shown in Eq.~\eqref{meanxcoherent}, our model predicts a third-harmonic component in $\langle x\rangle$, with a ratio between third- and first-harmonic amplitudes (third-harmonic distortion, $H_3$) equal to $H_3 = \epsilon \alpha^2/8$.  Using the definition of $\epsilon = l_k^2/(2 x_{\rm{zpm}}^2)$ an upper limit to the nonlocal length can be set in the form $l_k < 4\sqrt{H_3} \, x_{\rm{zpm}} /\alpha$. The bound on $l_k$ now also depends on $1/\alpha$ and can therefore be substantially lowered for high values of the coherent amplitude.

In order to prepare a quantum coherent state the system is first cooled down to its quantum ground state. As before, the measurement is then limited by the thermal decoherence time $\tau$. We remark that, with an intracavity power $P_{\rm{cav}}$, the oscillator is typically displaced from its equilibrium position, due to the radiation pressure,  by $x_0=(P_{\rm{cav}}/2c)/(m \omega^2)$. Once the cooling laser is turned off, this initial position determines a coherent state with amplitude $\alpha = x_0/\sqrt{2} x_{\rm{zpm}}$. The upper limit on the nonlocality scale can thus be written as $l_k < 4 \sqrt{2 H_3} \, x_{\rm{zpm}}^2 /x_0$, where $x_0$ is typically of the order of the cavity linewidth, i.e., $\sim10$ pm. As a consequence, it is not obvious to further excite the oscillator while cooling. On the other hand, one can conceive a coherent excitation with optical power $P_{\rm{exc}}$ just after the cooling stage, during a time interval $T_{\rm{exc}} \ll \tau$, and reach a displacement $x_0 = T_{\rm{exc}} \times (P_{\rm{exc}}/2c)/2 m\omega$. 

Using a low finesse optical cavity for this strong excitation pulse (this can be accomplished, e.g., by using the same cavity exploited for the optical cooling at a different wavelength), it is reasonable to achieve a $x_0$ of $\sim 1-10$ nm. The parameter $H_3$ could then be evaluated from the power spectral density of the signal monitoring the oscillator's position during a measurement period $\sim \tau$ following the excitation. We notice that a weak measurement (maybe with an optical signal seeing the cavity at low Finesse) is sufficient to detect a possible third harmonic signal that must be compared with the main coherent component, i.e. a sensitivity close to the quantum limit is not necessary here.

For the SiN or aluminum membranes mentioned before ($x_{\rm{zpf}} \simeq 1$ fm), one can aim to explore nonlocal scales down to $\sim 10^{-22}$ m. It is worth mentioning that an excitation yielding a coherent amplitude of several nanometers has already be applied to SiN membranes, and the experimental constraints on the third-harmonic distortion were similar to those that we are now considering \cite{bawaj}. The oscillator was in a thermal coherent state, but the results demonstrate that structural nonlinear effects are not a limit at this level. The use of the heavier oscillators discussed above could yield a further improvement of four orders of magnitude, to $\sim 10^{-26}$ m.

Besides the third-harmonic distortion, even the time-averaged variance is a useful indicator of possible nonlocal effects. To experimentally evaluate it one should first subtract, during the measurement period $\tau$, the coherent component (whose two parameters, amplitude and phase, can be extracted either from the complete decay or from the average over consecutive realizations) from the signal measuring $x$. The signal must then be time averaged along the interval $\tau$ and the mean square calculated over the result of repeated cycles. 

Our model predicts that the effects of nonlocality on Var$(x)$ are enhanced with respect to the ground state by the coherent amplitude $\alpha$. Taking the time average of Eq.~\eqref{varxcoherent} with $\alpha \gg 1$ we derive an upper limit to $\epsilon$ of the form $\epsilon <  \Delta_{\rm{meas}} / 3 \alpha^2$ and then $l_k < \sqrt{2 \Delta_{\rm{meas}}/3} \,x_{\rm{zpm}}/\alpha$, or $l_k < \sqrt{4 \Delta_{\rm{meas}}/3}\, x_{\rm{zpm}}^2/x_0 $. The expected achievable limits 
 on $l_k$ are roughly the same as those discussed for the harmonic distortion so that the evaluation of both indicators would provide a useful cross-check. However, we remark that in this case the dynamic range and the accuracy in the subtraction of the coherent component is a critical issue and could reduce the potential measurement sensitivity.

%

\section{Conclusions and Discussion}
\label{sec:concl}

We have shown that the nonrelativistic limit of an analytic nonlocal Klein-Gordon equation leads to a nonlocal
generalisation of the Schr\"odinger equation,~\eqref{nls}. We then constructed a phenomenological model in 
which the evolution of a 
single particle wavefunction in a harmonic oscillator potential is governed by such a nonlocal Schr\"odinger equation. This 
system is of particular interest to us because it is used to model the evolution of  quantum
optomechanical oscillators that are experimentally accessible. 
The introduction of the scale $\omega$ in the harmonic potential, allowed us to 
construct a small dimensionless parameter $\epsilon=m\omega l_k^2/\hbar$, with which 
we defined a perturbative expansion of the nonlocal differential operator $f(\Sc)$. 

We showed that a perturbative analysis of the nonlocal system~\eqref{seqn} leads to a sequence
of ordinary Schr\"odinger equations, order by order in $\epsilon$, 
with harmonic oscillator potential and in the presence of a source term. 
At every order $\epsilon^n$ the source term was shown to
depend on the solution to the problem at order $\epsilon^{n-1}$ for $n\ge1$.
We then solved the system of equations to first order in $\epsilon$ for perturbations
around a coherent state by using of the method of multiple scales. 
Having found that the resulting first order wavefunction failed to immediately lead to 
a well-defined probability measure, we normalised it such that a consistent probability 
measure could be defined. 

With this measure at hand we 
computed the expectation values and variances of observables $x$ and $p$. 
We found that the expectation values are unaltered for perturbations around the ground state ($\alpha=0$) 
but acquire a third harmonic for all $\alpha>0$ (except for the peculiar case $\alpha=1/\sqrt{6}$). 
Remarkably though, the state remains one of minimum uncertainty, i.e. Var($x$)Var($p$) $=1/4+O(\epsilon^2)$, 
for all $\alpha$, while undergoing a spontaneous periodic time-dependent squeezing in phase space. 

Finally, in Section \ref{sec:exp} we discussed how the prediction  
of both of these effects can be used to test the model experimentally. 
In particular we argued that, for the ground state, a comparison between the 
measured variance of $x$ and existing optomechanical macroscopic
oscillator experiments leads to bounds on $l_k$ of the order of $10^{-15}$~m. 
By imagining reasonable near future advances in these experiments we further argued that
bounds on $l_k$ of the order of $10^{-18}$~m could be achieved. 
This last number would provide an independent bound on nonlocality of the order
of the bound found using LHC data, with relatively inexpensive 
table-top experiments. 

Extending the analysis to comparisons between these experiments and the predicted  third
harmonic component in $\langle x\rangle$, in Section~\ref{evol} we argued that bounds of
the order of $l_k\sim 10^{-22}$~m can be achieved by looking at the evolution of coherent states. 
Improvements of four orders of 
magnitude are experimentally possible by using heavier oscillators, making constraints of order 
$l_k\sim 10^{-26}$~m possible in the near future. 
Similar bounds were then envisaged by making use of the oscillator's time-averaged variance, 
thus providing a potential cross-check of the previous analysis. 

Note that the effects related to a third harmonic in the evolution 
of $\langle x\rangle$ can only be used to cast constraints on $l_k$, since one can imagine 
similar effects being induced by the environment, and therefore only a lack of signal would be truly 
meaningful within this context. However our model does provide a ``smoking gun" of the nonlocal
evolution, namely the time dependent, periodic squeezing of the state in position and momenta, whose
magnitude grows with the coherent amplitude $\alpha$.
Indeed, there exist no other effects that we are aware of that could lead to such a spontaneous 
squeezing. In order to detect this effect though, one should adopt a different 
operative scheme with respect to the one discussed in Section~\ref{evol}. In particular, 
after the subtraction of the coherent component of the signal one should not average over 
the whole measurement interval $\tau$, but over time bins much shorter (say, $1/10$) than the 
oscillation period. The mean square would then be calculated over results belonging to several consecutive cycles 
for the same time bins. The time-dependent variance would thus be reconstructed. 
The sensitivity to nonlocal effects is expected to be similar to the one analysed for 
the time-averaged variance. 

Finally, let us now elaborate on the theoretical improvements that could (and should) 
be made to further strengthen our analysis. The model we constructed is phenomenological
in the sense that, although we were able to find a nonlocal Schr\"odinger equation for the
nonrelativistic field $\psi$, we did not show that this field could be consistently taken
to represent a one-particle quantum mechanical wavefunction satisfying the nonlocal 
Schr\"odinger equation~\eqref{nls}.

To fill this gap, one would need the Hamiltonian $H_{NL}$ of the nonlocal system, 
and use it to show that the one particle wave-function indeed satisfies the
equation
\be
i\hbar\fr{\partial\psi}{\partial t}=H_{NL}\psi,
\ee
where $H_{NL}$ would also contain higher order time derivatives.  
As well as providing a more direct link
between the underlying nonlocal QFT and the nonrelativistic quantum system, 
a Hamiltonian formulation would also
enable one to explicitly treat the oscillator as an open quantum system, via a generalised
kind of master equation. This description would allow for effects like decoherence and environmental noise
to be taken into account, and be better apt for investigating the effects 
of non-locality on the evolution of thermal coherent states, which are much easier to 
construct experimentally than pure coherent states and are therefore better suited 
to experimental comparison. 
It should also be noted that 
despite the fact that a preliminary analysis for thermal coherent states could be performed in the formalism
laid out in this paper,
\footnote{Recall that in the Glauber-Sudarshan $P$-representation the density matrix of a thermal coherent state is expressed 
in terms of pure coherent state projectors.} the Hamiltonian formalism would still be better suited
for this given that thermal coherent states are mixed.

The Hamiltonian analysis aside, recall that our computation of expectations values of observables 
forced us to normalise the wavefunction with its own norm in order to define a sensible probability density.
At the perturbative level one can check that the conserved charge~\eqref{prob} is {\em not} positive definite, 
but contains strongly suppressed negative regions. Thus, rigorously speaking, it cannot be interpreted 
as a probability density, something which points towards the fact that a positive definite charge can only
be obtained non-perturbatively. It is however possible to ignore these difficulties by using the charge, that is
conserved to first order $\epsilon$, in the computation of expectation values. When doing so the results for 
the expectation values and variances are qualitatively similar to the ones shown in this work and the 
phenomenologically relevant effect of spontaneous squeezing of coherent states persists.

To conclude, this analysis shows that opto-mechanical experiments have the potential to become a fundamental tool for high precision tests of quantum gravity-induced non-locality. 
Although achieving Planck scale sensitivities may not be strictly necessary in order to severely 
constrain certain quantum gravity scenarios, we believe that the rapid improvements 
of experimental techniques and instruments over recent years bode well for the possibility that
this scale may be closely approached in the next decade or so. 
It appears therefore that a new branch of quantum gravity phenomenology is about to begin, 
and we hope that the present work will further stimulate such turn of events.

\acknowledgments
A.B., D.M.T.B and S.L. wish to acknowledge the John Templeton Foundation for the supporting grant \#51876. The opinions expressed in this publication are those of the authors and do not necessarily reflect the views of the John Templeton Foundation. A.B. would like to thank Mauro Paternostro for stimulating discussions.
\appendix


\section{Multiple Scales Method}\label{app:multiple}
The method of multiple scales is needed for problems in which the solutions depend simultaneously on widely different scales. The method  introduces one or more new ‘slow’ time variables for each time scale of interest in the problem and subsequently treats these variables as if they are independent. 
To ensure a valid approximation to the solutions of our perturbation problem, we can use a simple two-scale expansion as the non local Schr\"odinger equation in dimensionless variables is characterized by the time evolution scale  $\omega=1$ and  the non locality scale $\varepsilon\ll1$. 
Here, the straightforward perturbative expansion in powers of $\epsilon$ leads to a nonuniform expansion where 
the perturbative ordering of the terms breaks down due to the presence of secular terms proportional to $\epsilon t$.  
The trick is to introduce a new variable $\hat{\tau}=\epsilon \hat{t}$, called the slow time because is not significant 
until $\hat{t} \sim 1/\epsilon$. Then the solution of the non local Schr\"odinger equation can be written as 
\begin{equation}
\psi(\hat{t},\hat{x})= \psi_{0}(\hat{t},\hat{\tau},\hat{x}) +\epsilon \psi_{1}(\hat{t},\hat{\tau},\hat{x})+ \dots 
\label{pex}
\end{equation}
Using the chain rule we have 

\begin{equation}
\fr{d}{d \hat{t}}  \psi = \fr{\del}{\del \hat{ t}} \psi_{0}  + \epsilon \left(\fr{\del}{\del \hat{\tau}} \psi_{0}  +\fr{\del}{\del \hat{t}} \psi_{1} \right) + \dots  \label{mse}
\end{equation}

Substituting \eqref{pex} into equation \eqref{nlse}, using \eqref{mse} and equating terms of order $\epsilon^0$ and $\epsilon^1$ 
gives
\begin{eqnarray}
\left(\hat{\mathcal{S}} - \frac{1}{2} \hat{x}^2\right) \psi_{0} &=& 0 \\  
 \left(\hat{\mathcal{S}} - \frac{1}{2} \hat{x}^2\right) \psi_{1}  &=& 2 a_2  \hat{\mathcal{S}}^2 \psi_{0}  + \fr{\del}{\del \hat{\tau}} \psi_{0} 
\end{eqnarray}
\begin{figure}[b]
\centering
\includegraphics[width=0.6\textwidth]{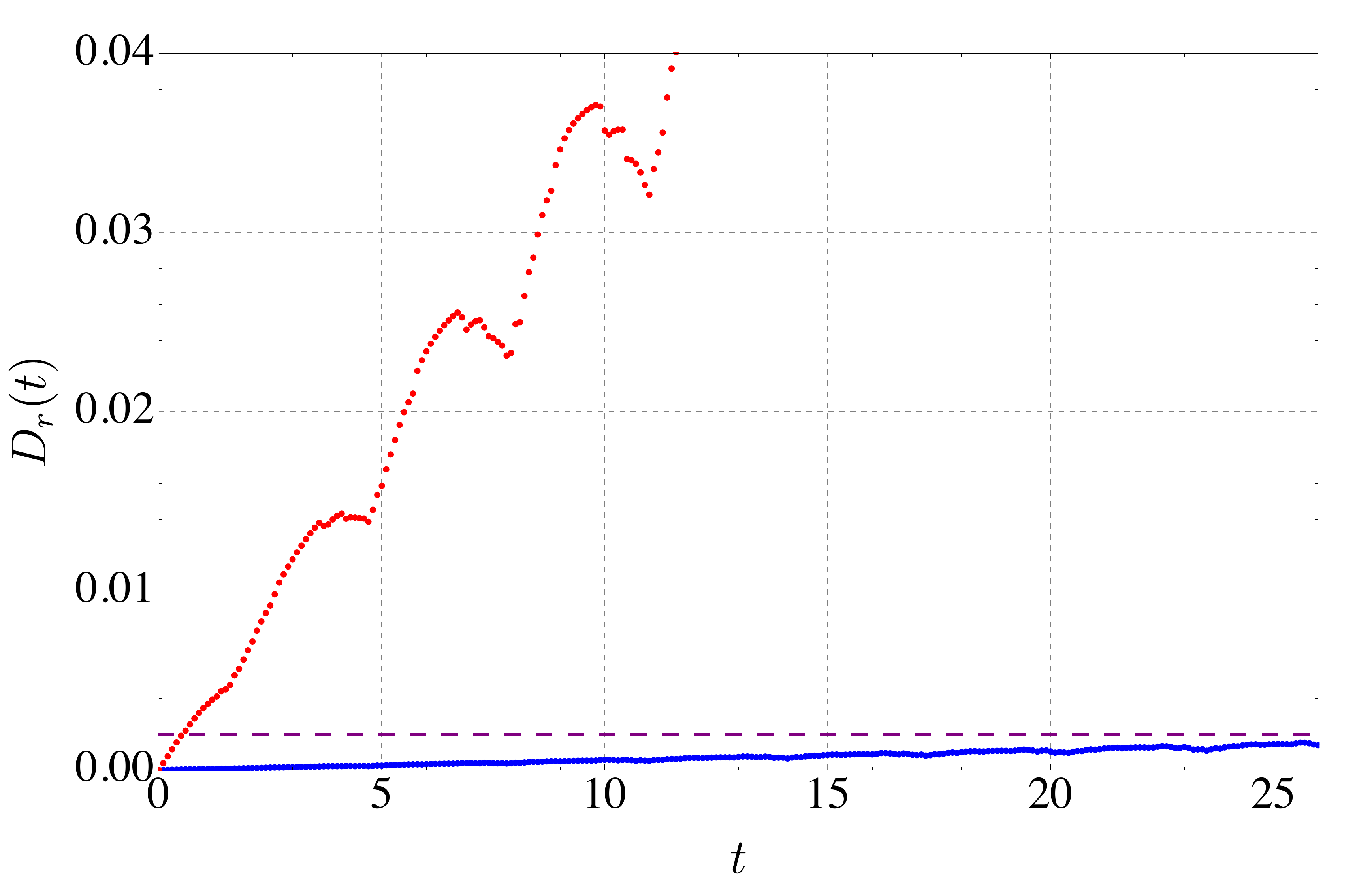} 
\caption{Time dependence of the relative distance $D_r(\hat{t}_k)\equiv D(\hat{t}_k)/ \max_j\left\{ |\psi(\hat{t}_k,\hat{x}_j)|\right\}$ among perturbed solutions of the non-local Schr\"odinger equation (color online). For the numerical solution, we fix $\alpha=1$ and $a_2 \epsilon = 2\times 10^{-3}$.  The horizontal purple line is the relative distance threshold $2 \times 10^{-3}$.  Blue dots represent the relative distance between numerical and analytical solutions. Red dots represent the relative distance between numerical and analytical solutions without dropping secular terms. }\label{norm}
\end{figure}
To ensure that there are no secular terms in our ansatz solution  \eqref{eqanz}  for coherent states,  the terms proportional to $\hat{t}$ in $\psi_1(\hat{t},\hat{x})$ are forced to be 0 by assuming 
\begin{equation}
\psi_0(\hat{t},\hat{\tau},\hat{x})= \psi_{0}(\hat{t},\hat{x}) \exp[ \hat{\tau} f(\hat{t},\hat{x}) \hat{t} ] \ ,
\label{eqms}
\end{equation}
where $f(\hat{t},\hat{x})$ is a suitable polynomial in $\hat{x}$ with time dependent coefficients.  
To confirm the reliability of our solution method in Section~\ref{nloc-Sch}, we have checked that there are no solutions growing in time as 
fast as $\epsilon \hat{t}$ by numerically solving the non local Schr\"odinger equation. To this end, we  solved 
the equation  
$$\left(\hat{\mathcal{S}} - 2 a_2 \epsilon \hat{\mathcal{S}}^2  - \hat{x}^2/2 \right) \psi(\hat{t},\hat{x})=0 $$ 
in the rectangular domain $  [-6 \leq x \leq 6] \times [0\leq t \leq 25] $ of the space-time plane,   with $2 a_2 \epsilon=10^{-3}$ and periodic boundary conditions in space. In addition, we set the initial conditions
\begin{align*}
& \psi(0,\hat{x})= \frac{1}{\pi^{1/4} } \ \exp\left[\sqrt{2} \hat{x} -\frac{\hat{x}^2}{2}-1\right] \\
& \left.\frac{d}{d\hat{t}}\psi(\hat{t},\hat{x})\right|_{\hat{t}=0}=-\frac{i e^{-\frac{\hat{x}^2}{2}+\sqrt{2} \hat{x}-1} \left(2 \sqrt{2} \hat{x}-1\right)}{2 \pi^{1/4} }\ ,
\end{align*} 
representing the  $\alpha=1$ coherent state. 
The numerical solution was calculated using the implicit Euler method of the partial differential equation solver provided by Mathematica. To quantify numerical errors in the discrete space and time domains, we introduce the Chebyshev distance between solutions $\psi_1$ and $\psi_2$  as 
\begin{equation}
D(\hat{t}_k)\equiv \max_j \left\{ |\psi_1(\hat{t}_k,\hat{x}_j) - \psi_2(\hat{t}_k,\hat{x}_j)| \right\}. 
\label{eqcheb}  
\end{equation}
To calculate $D(\hat{t}_k)$, we set the space mesh size to $10^{-2}$ and the time mesh size to $10^{-1}$.

Fig.~\ref{norm} shows the plots of the relative maximum distances between the numerical and analytical solutions  either removing or keeping terms $\epsilon \hat{t}$. These plots clearly show that secular terms in the polynomial coefficients $c_0(\hat{t})$, $c_1(\hat{t})$, $c_2(\hat{t})$, $c_3(\hat{t})$,  and $c_4(\hat{t})$ of $\psi_1$ have been properly discarded. We stress that the small mismatch in Fig. \ref{norm} between numerical and analytical solutions is due to the accumulation of numerical errors at large time as it does not grow as fast as $\epsilon \hat{t}$. 
As a final remark we also point out that there is good agreement between mean and variance of position and momentum in 
 eqs. (\ref{meanxcoherent}) - (\ref{varpcoherent}) evaluated with $\alpha=1 $, $2 a_2 \epsilon=10^{-3}$, and the same quantities estimated by means of the numerical solution.

\bibliographystyle{apsrev}
\bibliography{refs.bib}

\end{document}